\documentclass[superscriptaddress,preprintnumbers,amsmath,amssymb,twocolumn,floats]{revtex4-2}
\usepackage{txfonts}
\usepackage{amssymb}
\usepackage{amsfonts}
\usepackage{amsmath}
\usepackage{graphicx}
\usepackage{epstopdf}
\usepackage{color}
\begin{document}

\title{Anomalous thermoelectric effects and quantum oscillations in the kagome metal CsV$_3$Sb$_5$}

\begin{abstract}
The kagome metal compounds $A$V$_3$Sb$_5$ ($A$ = K, Rb, and Cs) feature a wealth of phenomena including nontrivial band topology, charge density wave (CDW), and superconductivity. One intriguing property is the time-reversal symmetry breaking in the CDW state without local moments, which leads to anomalous transport responses. Here, we report the investigation of magneto-thermoelectric effects on high-quality CsV$_3$Sb$_5$ single crystals. A large anomalous Nernst effect is observed at temperatures below 30 K. Multiple Fermi surfaces with small effective masses are revealed by quantum oscillations in Nernst and Seebeck signals under high magnetic field. Furthermore, we find an unknown frequency, and attribute it to the magnetic breakdown across two smaller Fermi surfaces. A gap around 20 meV can be resolved from the breakdown threshold field, which we propose to be introduced by the CDW. These results shed new light on the CDW-related phenomena, particularly in $A$V$_3$Sb$_5$ compounds. 

\end{abstract}

\author{Dong Chen}
\email{Dong.Chen@cpfs.mpg.de}
\affiliation{Max Planck Institute for Chemical Physics of Solids, 01187 Dresden, Germany}
\affiliation{College of Physics, Qingdao University, Qingdao 266071, China}

\author{Bin He}
\affiliation{Max Planck Institute for Chemical Physics of Solids, 01187 Dresden, Germany}

\author{Mengyu Yao}
\affiliation{Max Planck Institute for Chemical Physics of Solids, 01187 Dresden, Germany}

\author{Yu Pan}
\affiliation{Max Planck Institute for Chemical Physics of Solids, 01187 Dresden, Germany}

\author{Haicheng Lin}
\affiliation{Max Planck Institute for Chemical Physics of Solids, 01187 Dresden, Germany}

\author{Walter Schnelle}
\affiliation{Max Planck Institute for Chemical Physics of Solids, 01187 Dresden, Germany}

\author{Yan Sun}
\affiliation{Max Planck Institute for Chemical Physics of Solids, 01187 Dresden, Germany}

\author{Johannes Gooth}
\affiliation{Max Planck Institute for Chemical Physics of Solids, 01187 Dresden, Germany}

\author{Louis Taillefer}
\affiliation{Institut Quantique, D\'{e}partement de Physique and RQMP, Universit\'{e} de Sherbrooke, Sherbrooke, Qu\'{e}bec J1K 2R1, Canada}
\affiliation{Canadian Institute for Advanced Research, Toronto, Ontario M5G 1M1, Canada}

\author{Claudia Felser}
\email{Claudia.Felser@cpfs.mpg.de}
\affiliation{Max Planck Institute for Chemical Physics of Solids, 01187 Dresden, Germany}
\affiliation{Canadian Institute for Advanced Research, Toronto, Ontario M5G 1M1, Canada}

\maketitle

Condensed matter systems with kagome lattices have attracted significant interest owing to their rich physics. With the special two-dimensional corner-sharing triangular network, the electronic structure of the kagome lattice holds Dirac cones, flat bands, and enhanced correlation \cite{kagomeTB,SLnature}. Further inclusion of other collective orders in metallic kagome materials can give rise to more exotic quantum states and phenomena \cite{Fe3Sn2,TbMn6Sn6,Co3Sn2S2,Co3Sn2S2Lei,CSSANE,Mn3Sn,Mn3SnANE,Mn3Ge}. Recently, a new kagome metal family $A$V$_3$Sb$_5$ ($A$ = K, Rb, and Cs) was discovered, hosting a superconducting ground state with the critical temperature ($T\rm_c$) ranging from 0.93 K for KV$_3$Sb$_5$ to 2.5 K for CsV$_3$Sb$_5$, as well as a charge-density-wave (CDW) transition at $T\rm_{CDW}$ = 80, 103, and 94 K, respectively \cite{AV3Sb5,KVSsupc,RVSsupc,CVSsupc}. Band structure calculations revealed multiple van Hove singularities, Dirac cones and the nontrivial $\mathbb{Z}_2$ topology \cite{CVSsupc,CVSDFTYan}. The unusual band structure modulated by CDW is relevant for various interesting properties, such as unconventional superconductivity, chiral charge order, and time-reversal symmetry breaking (TRSB) without local moments \cite{CVSsupc,CVSDFTYan,KVSCDW,CVSSTM1,CVSSTM2,CVSSTMXHChen,KVSAHE,CVSAHE,CVSuSR1,KVSuSR,CVSuSR2}.

Generally, in a band structure with Dirac cones close to the Fermi level, the TRSB can open an energy gap at the Dirac points, thereby generating a large Berry curvature \cite{WeylQi}. The Berry curvature acts like a magnetic field in the momentum space, and lead to the anomalous Hall effect (AHE) and anomalous Nernst effect (ANE), even without ferromagnetism \cite{BerryRMP}. According to the Mott relation, the thermoelectric effects are proportional to the energy derivative of the electric conductivities at the Fermi level \cite{zimanE&P}. The thermoelectric effects are more sensitive to the Berry curvature near the Fermi level than the respective electric conductivities. The large ANE induced by the Berry curvature has been observed in Cd$_3$As$_2$ and ZrTe$_5$ \cite{Cd3As2ANE,ZrTe5ANE}. Furthermore, the quantum oscillations in the thermoelectric coefficients are typically stronger compared to the Shubnikov-de Haas (SdH) oscillations, which can provide relevant information on the band structure \cite{WTeBehnia}.

In this letter, we report a study of magneto-thermoelectric properties in high-quality single crystals of CsV$_3$Sb$_5$. We observe large ANE below temperature of 30 K. The Seebeck coefficient shows a mustache shape in low field, as a consequence of the multiband effect. Significant quantum oscillations are present in both Seebeck and Nernst signals, revealing multiple Fermi surfaces (FSs) with small effective masses. Besides the frequencies commonly found in the Shubnikov-de Haas (SdH) oscillations, the thermoelectric signals show an additional frequency under high magnetic field. We attribute this frequency to the magnetic breakdown across two smaller FSs that are split from a single Dirac band by a CDW gap. These results can help us to have a deeper understanding on the TRSB and the electronic structure reconstruction induced by CDW in this system.

High-quality single crystals of CsV$_3$Sb$_5$ were grown using the self-flux method \cite{CVSsupc}. The transport properties were measured in a Quantum Design physical properties measurement system. The Seebeck and Nernst signals were measured by a self-built one-heater-two-thermometers setup with $H \parallel c$ and $-\nabla T \parallel a$ crystallographic directions. We report the thermoelectric results for two samples, \#1 and \#2. The temperature difference on sample \#1 was measured with Cernox thermometers from 2 to 25 K, while on sample \#2 it was determined with type-E thermocouples from 10 to 300 K.

\begin{figure}
	\includegraphics[width=8.6cm]{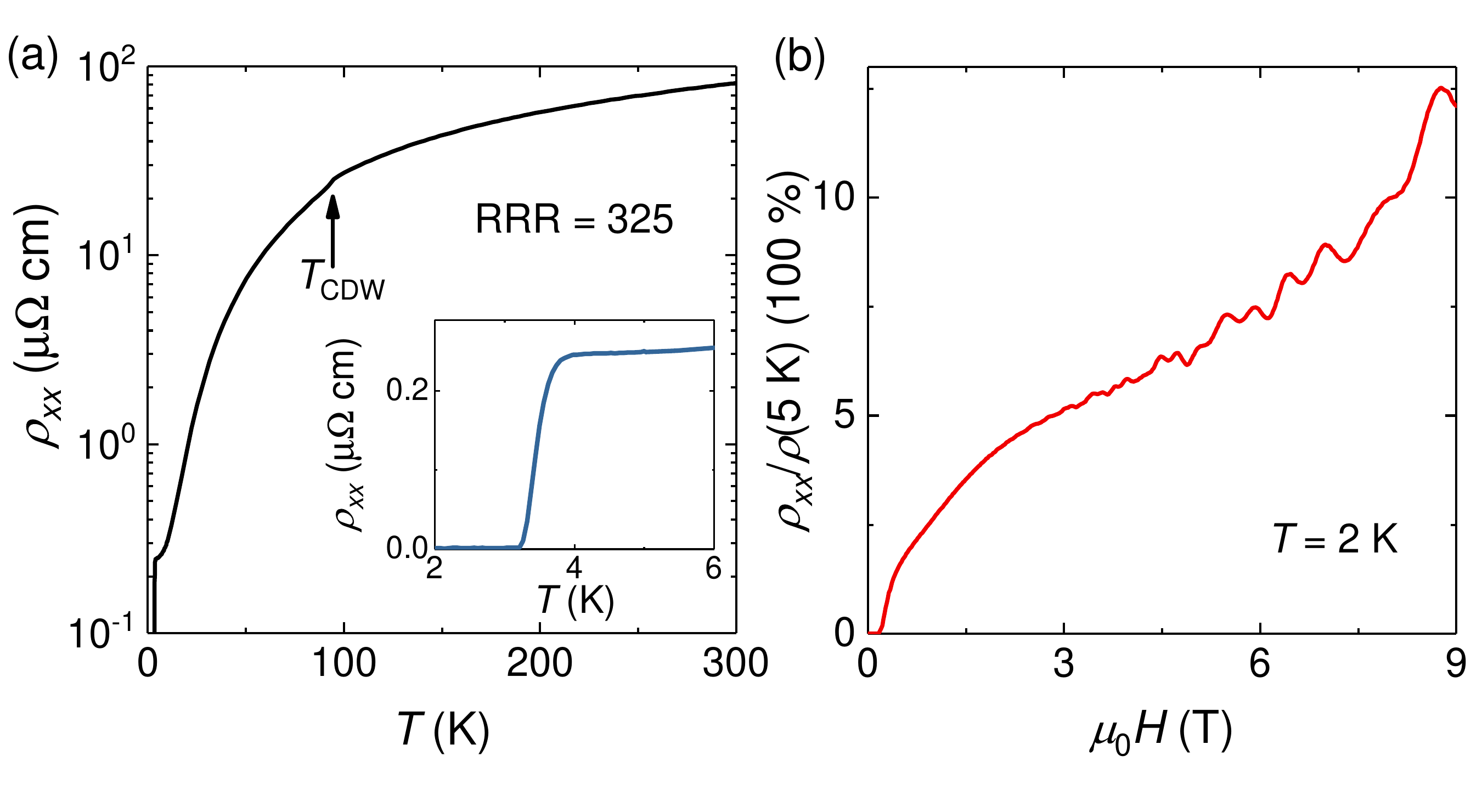}
	\caption{
		\label{f1}(Color online)
		(a) Temperature dependence of the resistivity of a typical sample with the RRR as high as 325. The kink at 94 K corresponds to the CDW transition. The inset shows the low-temperature resistivity with a sharp superconducting transition at 3.5 K.
		(b) Resistivity at 2 K divided by the normal-state resistivity $\rho$(5 K) as a function of the magnetic field. 
		The high RRR and magnetoresistance, sharp superconducting transition, and intense quantum oscillations suggest the high quality of the samples.  
	}
\end{figure}

Figure \ref{f1} shows the resistivity ($\rho_{xx}$) of a typical crystal as the functions of temperature and magnetic field. The resistivity displays a kink due to the CDW transition at 94 K, and a superconducting transition at 3.5 K. All the behaviors are similar to previous reports \cite{CVSsupc,CVSAHE}, except for the extremely large residual resistivity ratio RRR = $\rho$(300 K)/$\rho$(5 K) of 325 [Fig. \ref{f1}(a)]. Also, the sample shows a much larger magnetoresistance and evident SdH oscillations as shown in Fig. \ref{f1}(b). The magnetoresistance at 2 K referenced to $\rho$(5 K) exceeds 1000\% at 9 T. All the results indicate the low defects density that guarantees the long mean free path of carriers in cryogenic temperature. These features can provide the preferable condition for large ANE and magnetic breakdown, which are discussed in the following parts.

\begin{figure}
	\includegraphics[width=8.6cm]{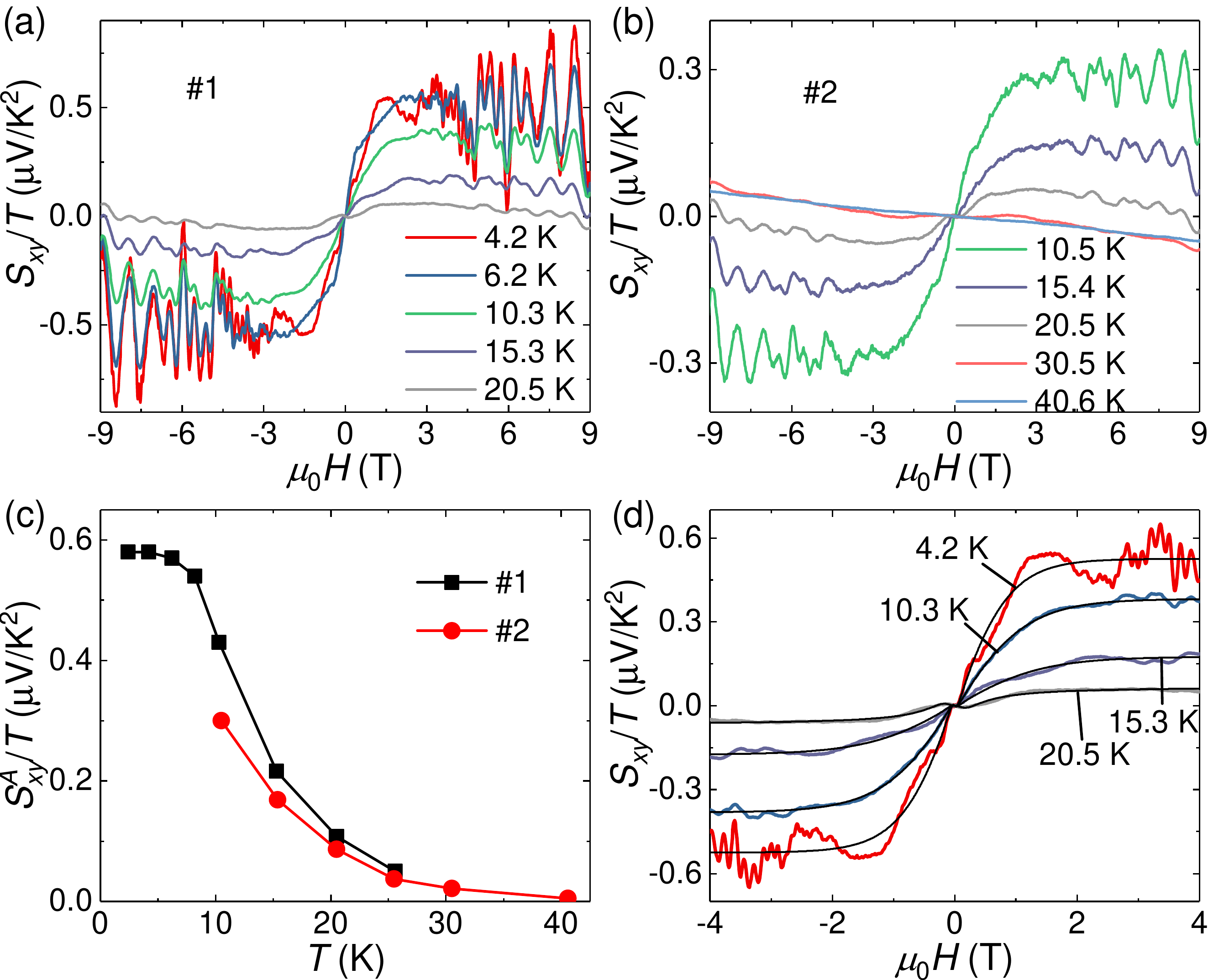}
	\caption
	{
		\label{f2}(Color online) 
		(a, b) Magnetic field dependence of the Nernst signal $S_{xy}/T$ at various temperatures for samples \#1 and \#2, respectively. 
		(c) Anomalous Nerst signals of the two samples as a function of the temperature. 
		(d) Fitting of the Nernst signal with the empirical expression (1) at selected temperatures. The black lines indicate the fitting curves.
	}
\end{figure}

Figures \ref{f2}(a) and (b) show the magnetic field dependence of the Nernst signal $S_{xy}/T$ for samples \#1 and \#2 at selected temperatures, respectively. In the semiclassical one-band theory, the Nernst thermopower $S_{xy}$ evolves with magnetic field as $S_{xy}=S_0{\mu B}/[1+(\mu B)^2]$, where $\mu$ is the carrier mobility. It has a peak at $B=1/\mu$ and tends to zero under higher field \cite{PbSe}. The Nernst signal of both samples \#1 and \#2 show a weak peak below 1 T, and the peak field value increases with temperature. The low peak field underlines the high mobility of the charge carriers. At low temperatures, the Nernst signals display intense quantum oscillations, and tend to a nonzero constant in high-field region, which is an obvious anomalous component. As the temperature increases, the anomalous component gradually vanishes. At higher temperatures, a linear behavior with a negative slope becomes noticeable, which is caused by the multiband effect \cite{NbPNernst,CVSTE}.

To further reveal the origin of the ANE, we extract the anomalous Nernst component by the linear extrapolation of $S_{xy}(B)$ to zero field, as shown in Fig. \ref{f2}(c). Although the anomalous component for sample \#1 is larger than that of sample \#2, they have a similar temperature dependence in the overlapping range. The ANE component has almost no temperature dependence at low temperatures, and rapidly decreases at higher temperatures until about 30 K. The Nernst curves can also be described by an empirical expression \cite{Cd3As2ANE}: 
\begin{equation}
S_{xy}(B)=S^N_{xy}\frac{\mu B}{1+(\mu B)^2}+S^A_{xy}\rm{tanh}\left( \frac{\textit{B}}{\textit{B}_0}\right),
\end{equation}
where $S^N_{xy}$ and $S^A_{xy}$ are the ordinary and anomalous Nernst signal amplitudes, respectively. $B_0$ is the saturation field of the anomalous component. This expression fits well with the Nernst signals for low temperatures, and some examples of the fits are shown in Fig. \ref{f2}(d). They result in a very high mobility of $\mu \sim$ 10$^5$ cm$^2$V$^{-1}$s$^{-1}$ for sample \#1 and an order of magnitude lower for sample \#2. The mobility difference can also be perceived by comparing the Nernst signals of the two samples at same temperature, where the quantum oscillations of sample \#1 are obviously greater than those of sample \#2. These features suggest the enhancement effect of the mobility to the ANE \cite{Cd3As2ANE}.

\begin{figure}
	\includegraphics[width=8.6cm]{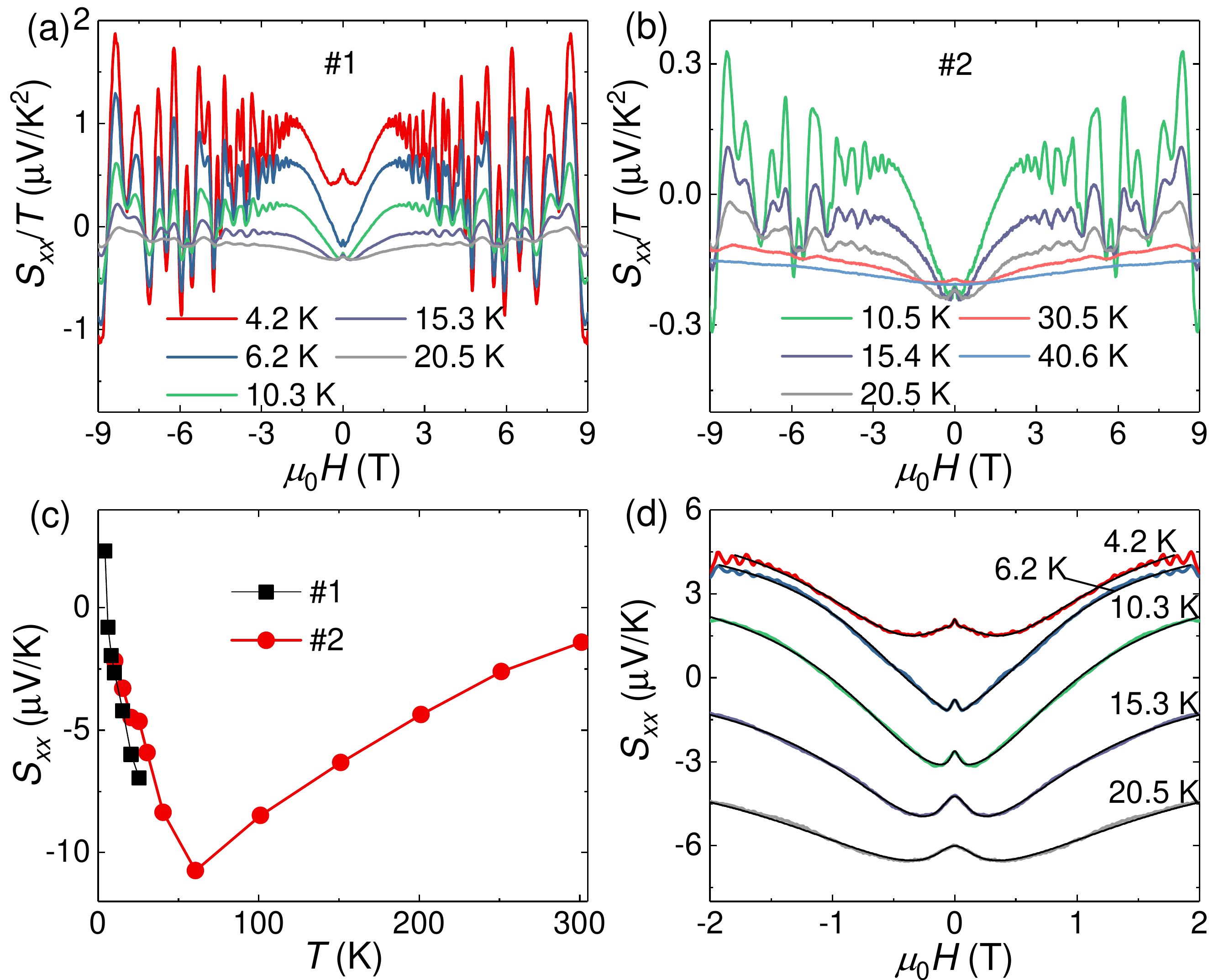}
	\caption
	{
		\label{f3}(Color online) 
		(a, b) Magnetic field dependence of the Seebeck signals for sample \#1 and \#2, respectively.
		(c) Temperature dependence of the Seebeck coefficient at the zero field for both samples. 
		(d) Seebeck signal of sample \#1 with the fitting lines of the two-band expression (2).
	}
\end{figure}

Figures \ref{f3}(a) and (b) show the Seebeck coefficient $S_{xx}$ for both samples at selected temperatures. Also, the Seebeck signals of the two samples have similar magnetic field dependence and close values. The temperature dependence of the zero-field $S_{xx}$ for both samples are shown in Fig. \ref{f3}(c). It is negative for $T >$ 6 K, indicating the dominant electron carriers, but becomes positive for lower temperatures, proving the existence of two types of carriers. Moreover, for temperatures below $\approx 20$ K the Seebeck signal has a mustache shaped profile around zero field, which cannot be explained by the conventional one-band model \cite{PbSe}. Considering the multiband nature of the system, we use the modified expression:
\begin{equation}
   S_{xx}(B)=S_1\frac{1}{1+(\mu_1 B)^2}+S_2\frac{1}{1+(\mu_2 B)^2}+S_\infty\frac{(\mu^\prime B)^2}{1+(\mu^\prime B)^2},
\end{equation}
where $S_1$ ($S_2$) and $\mu_1$ ($\mu_2$) are the zero-field Seebeck coefficients and mobility of the first (second) carrier, respectively, and $S_\infty$ is the limiting value when $B\rightarrow\infty$. The expression can well fit the Seebeck signals, as shown in Fig. \ref{f3}(d). 

\begin{figure}
	\includegraphics[width=8.6cm]{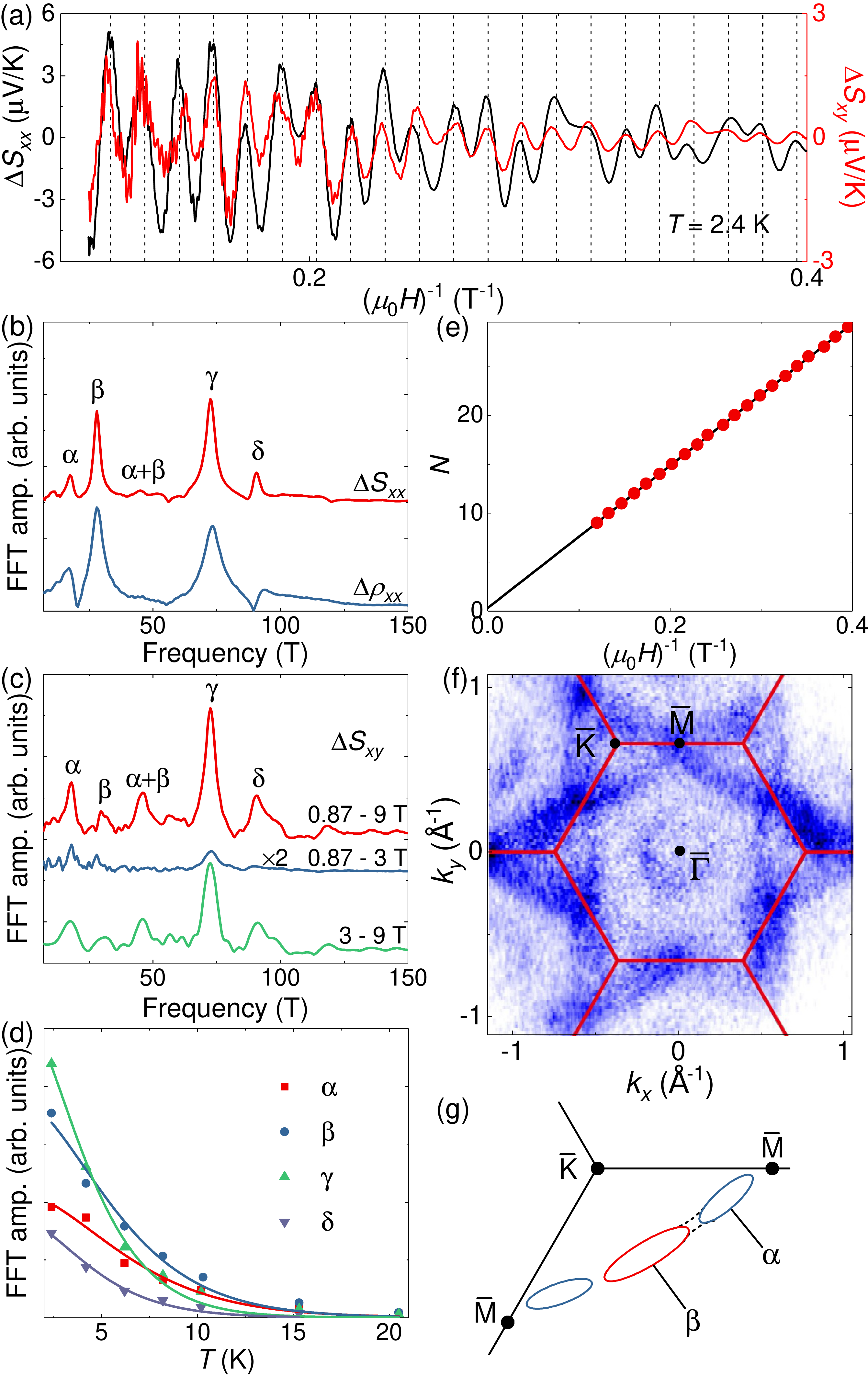}
	\caption
	{
		\label{f4}(Color online) 
		(a) Oscillation parts of both $S_{xx}$ and $S_{xy}$ at 2.4 K. A primary frequency of $F_\gamma$ = 72 T can be identified and the peaks are marked by dashes lines.
		(b) FFT spectrum of the Seebeck and SdH oscillations with four peaks are labeled. The highest peak $\gamma$ corresponds to the primary frequency shown in (a).
		(c) FFT spectrum of the Nernst oscillations obtained from different magnetic ranges. A fifth peak with frequency close to $F_\alpha + F_\beta$ appears but is absent below 3 T.
		(d) Amplitudes of the four main peaks as a function of the temperature, with the solid lines representing the LK fitting. The effective masses obtained from $\Delta S_{xx}$ and $\Delta S_{xy}$ have very close values.
		(e) Index plot obtained from the primary oscillation of $\Delta S_{xx}$ shown in (a). 
		(f) FS mapping measured by the ARPES at 33 K with the Brillouin zone and high-symmetry points superposed.
		(g) Possible sketch of FSs $\alpha$ and $\beta$, with a breakdown path marked by the dashed lines.
	}
\end{figure}

The quantum oscillations can provide more information on the electronic structure. Figure \ref{f4}(a) shows the oscillatory parts $\Delta S_{xx}$ and $\Delta S_{xy}$ at 2.4 K. Although composed by multiple frequencies, the oscillations have one primary frequency for both $\Delta S_{xx}$ and $\Delta S_{xy}$ as marked by the dashed lines. As shown in Fig. \ref{f4}(b), the fast Fourier transformation (FFT) of $\Delta S_{xx}$ reveals four main frequencies, which are $F_\alpha$ = 18 T, $F_\beta$ = 28 T, $F_\gamma$ = 72 T, and $F_\delta$ = 91 T, consistent with those in the SdH oscillations \cite{CVSAHE,CVSFSmapping}. From the Onsager relation $F = (\hbar/2\pi e)S\rm_F$ \cite{shoenberg}, these low frequencies correspond to four small FSs. Moreover, there is an additional significant frequency at approximately 46 T in the FFT of Nernst oscillations, and this frequency is less notable in the Seebeck and SdH oscillations [Fig. \ref{f4}(c)]. This additional frequency is not a harmonic one of other frequencies but roughly equals to the sum of $F_\alpha$ and $F_\beta$. Besides, it only occurs in the field region $>$ 3 T, as shown in Fig. \ref{f4}(c). We attribute it to the magnetic breakdown across the orbits $\alpha$ and $\beta$. This quantum tunneling effect suggests their adjacent positions in the Brillouin zone \cite{shoenberg}.

The Lifshitz-Kosevitch (LK) theory describes the evolution of the quantum oscillations with the temperature and magnetic field \cite{shoenberg}, where the cyclotron effective mass and Dingle temperature are involved. For the oscillations in the thermoelectric coefficients, we fit the temperature dependence of the amplitudes of the oscillations using the following expression \cite{Bi2Se3Behnia}:
\begin{equation}
\frac{A}{T}\propto\frac{\lambda}{\rm sinh(\lambda)},
\end{equation}
where $A$ is the amplitude of $\Delta S_{xx}$ or $\Delta S_{xy}$,  $\lambda=2\pi^2k_Bm^*T/e\hbar B$, and $m^*$ is the cyclotron effective mass. For $B$, we use the average of the field range of oscillations, $1/B=1/B_1+1/B_2$. Figure \ref{f4}(d) shows the selected fitting results. Both $\Delta S_{xx}$ and $\Delta S_{xy}$ give consistent effective masses: $m^*_\alpha = 0.039$ $m_0$,  $m^*_\beta = 0.043$ $m_0$, $m^*_\gamma = 0.058$ $m_0$, and $m^*_\delta = 0.054$ $m_0$. These light effective masses are close to the ones obtained from SdH oscillations \cite{CVSAHE}.

Figure \ref{f4}(e) shows the index plot of the primary oscillation shown in Fig. \ref{f4}(a), and the ($\mu_0 H)^{-1}$ values correspond to the maxima of the $\Delta S_{xx}$ or $\Delta S_{xy}$. The slope of the index plot gives a frequency of 72 T, which is consistent with the value of $F_\gamma$ obtained by the FFT. Because $S_{xx}$ and $S_{xy}$ are the diagonal and off-diagonal terms of the tensor $S$, respectively, the maxima in $\Delta S_{xy}$ typically have a 1/4 phase shift relative to $\Delta S_{xx}$ \cite{PbSe}. However, there is no phase shift between $\Delta S_{xx}$ and $\Delta S_{xy}$ in CsV$_3$Sb$_5$, at least at the frequency $F_\gamma$. The reason remains unclear and requires further investigation.

Understanding the band structure in the CDW state of the $A$V$_3$Sb$_5$ compounds is a key topic. Although density-functional-theory calculations can well explain the band structure observed in angle-resolved photoemission spectroscopy (ARPES) experiments \cite{CVSsupc}, the four oscillation frequencies in the transport measurements remain poorly understood \cite{CVSFSmapping,CVSLei}. It suggests the delicate effect of the CDW modulation to the band structure. The electronic structure of undistorted CsV$_3$Sb$_5$ has three types of bands in the vicinity of the Fermi level: a parabolic electronic band near the $\overline{\Gamma}$ point, multiple Dirac bands around the $\overline{K}$ points, and saddle points or van Hove singularities at the $\overline{M}$ points \cite{CVSsupc,CVSDFTYan}, as shown in Fig. \ref{f4}(f). The CDW transition in the $A$V$_3$Sb$_5$ system is commonly believed to be driven by the Peierls instability related to FS nesting \cite{CVSDFTYan}. The direct consequence of this instability is the gap opening on the FSs, which has been demonstrated by several experimental techniques \cite{CVSSTM1,CVSSTM2,CVSSTMXHChen,CVSgapWen,CVSgapARPES1,CVSgapARPES3,CVSgapARPES2,KVSgapARPES,RVSgapARPES}. Moreover, the CDW gap has a strong momentum dependence along the FSs of the Dirac bands \cite{CVSgapARPES2,KVSgapARPES}. The small effective masses obtained from our quantum oscillations indicate that these low-frequency oscillations originate from the Dirac bands. The FSs $\alpha$ and $\beta$ may originate from a single band split by the CDW gap. Figure \ref{f4}(g) shows a possible sketch of the FSs $\alpha$ and $\beta$, together with a breakdown path. The gap ($\epsilon_g$) between them can be related to the threshold field of magnetic breakdown ($B^* \sim$ 3 T) by $\hbar\omega_c \geq \epsilon^2_g/E\rm_F$, where $\omega_c = eB^*/m^*$ \cite{shoenberg}. With $E\rm_F \sim$ 50 - 80 meV given by the oscillations, the gap $\epsilon_g$ can be estimated less than $\sim$ 20 meV. This value is consistent with the CDW gap obtained from the spectroscopy experiments.

In the CDW state of $A$V$_3$Sb$_5$ compounds, an unconventional chiral charge order was discovered \cite{KVSCDW,CVSCDW}, and an accompanied chiral flux phase was proposed to account for the TRSB \cite{AVSCDW,AVSCDW2}. For CsV$_3$Sb$_5$, the TRSB has been confirmed by a $\mu$SR experiment \cite{CVSuSR2}, along with the AHE and our ANE \cite{CVSAHE,CVSAHEgate}. Intriguingly, the TRSB signal in the $\mu$SR experiment, the AHE, and the ANE appear below different temperatures. A local field within the kagome plane is detected by the $\mu$SR below 70 K ($\textless$ $T\rm_{CDW}$), and an additional perpendicular component appears below 30 K. However, the AHE starts at $T\rm_{CDW}$, whereas in our data the ANE becomes obvious below 30 K. The different temperature regimes where AHE and ANE are noticed may be due to their different sensitivities in this system. The ANE appears at the close temperature as the out-of-plane field in the $\mu$SR data, suggesting a possible relationship between the ANE and a hidden chiral flux phase.

In summary, we conducted a systematic investigation on the magneto-thermoelectric effect of the kagome metal CsV$_3$Sb$_5$. The Nernst signal shows a large anomalous component below 30 K, which suggests the time-reversal symmetry breaking may be accompanied by a hidden chiral flux phase appearing at the same temperature. The quantum oscillations in the thermoelectric coefficients reveal multiple Fermi surfaces with small effective masses. Two small Fermi surfaces that are split from a single Dirac band by a charge density wave gap are indicated by the magnetic breakdown effect. Our results are significant for further study on the chiral flux phase, fine electronic structure, and novel superconductivity in the charge density wave state of kagome metal.

This work was supported by the European Research Council Advanced Grant (No. 742068) ``TOPMAT", the Deutsche Forschungsgemeinschaft (Project-ID No. 258499086) ``SFB 1143", and the DFG through the W\"urzburg-Dresden Cluster of Excellence on Complexity and Topology in Quantum Matter ct.qmat (EXC 2147, Project-ID No. 39085490).

\bibliography{CsV3Sb5ref}

\begin{thebibliography}{47}
\expandafter\ifx\csname natexlab\endcsname\relax\def\natexlab#1{#1}\fi
\expandafter\ifx\csname bibnamefont\endcsname\relax
  \def\bibnamefont#1{#1}\fi
\expandafter\ifx\csname bibfnamefont\endcsname\relax
  \def\bibfnamefont#1{#1}\fi
\expandafter\ifx\csname citenamefont\endcsname\relax
  \def\citenamefont#1{#1}\fi
\expandafter\ifx\csname url\endcsname\relax
  \def\url#1{\texttt{#1}}\fi
\expandafter\ifx\csname urlprefix\endcsname\relax\def\urlprefix{URL }\fi
\providecommand{\bibinfo}[2]{#2}
\providecommand{\eprint}[2][]{\url{#2}}

\bibitem[{\citenamefont{Ohgushi et~al.}(2000)\citenamefont{Ohgushi, Murakami,
  and Nagaosa}}]{kagomeTB}
\bibinfo{author}{\bibfnamefont{K.}~\bibnamefont{Ohgushi}},
  \bibinfo{author}{\bibfnamefont{S.}~\bibnamefont{Murakami}}, \bibnamefont{and}
  \bibinfo{author}{\bibfnamefont{N.}~\bibnamefont{Nagaosa}},
  \bibinfo{journal}{Phys. Rev. B} \textbf{\bibinfo{volume}{62}},
  \bibinfo{pages}{R6065(R)} (\bibinfo{year}{2000}).

\bibitem[{\citenamefont{Balents}(2010)}]{SLnature}
\bibinfo{author}{\bibfnamefont{L.}~\bibnamefont{Balents}},
  \bibinfo{journal}{Nature} \textbf{\bibinfo{volume}{464}},
  \bibinfo{pages}{199} (\bibinfo{year}{2010}).

\bibitem[{\citenamefont{Ye et~al.}(2018)\citenamefont{Ye, Kang, Liu, Von~Cube,
  Wicker, Suzuki, Jozwiak, Bostwick, Rotenberg, Bell et~al.}}]{Fe3Sn2}
\bibinfo{author}{\bibfnamefont{L.}~\bibnamefont{Ye}},
  \bibinfo{author}{\bibfnamefont{M.}~\bibnamefont{Kang}},
  \bibinfo{author}{\bibfnamefont{J.}~\bibnamefont{Liu}},
  \bibinfo{author}{\bibfnamefont{F.}~\bibnamefont{Von~Cube}},
  \bibinfo{author}{\bibfnamefont{C.~R.} \bibnamefont{Wicker}},
  \bibinfo{author}{\bibfnamefont{T.}~\bibnamefont{Suzuki}},
  \bibinfo{author}{\bibfnamefont{C.}~\bibnamefont{Jozwiak}},
  \bibinfo{author}{\bibfnamefont{A.}~\bibnamefont{Bostwick}},
  \bibinfo{author}{\bibfnamefont{E.}~\bibnamefont{Rotenberg}},
  \bibinfo{author}{\bibfnamefont{D.~C.} \bibnamefont{Bell}},
  \bibnamefont{et~al.}, \bibinfo{journal}{Nature}
  \textbf{\bibinfo{volume}{555}}, \bibinfo{pages}{638} (\bibinfo{year}{2018}).

\bibitem[{\citenamefont{Yin et~al.}(2020)\citenamefont{Yin, Ma, Cochran, Xu,
  Zhang, Tien, Shumiya, Cheng, Jiang, Lian et~al.}}]{TbMn6Sn6}
\bibinfo{author}{\bibfnamefont{J.-X.} \bibnamefont{Yin}},
  \bibinfo{author}{\bibfnamefont{W.}~\bibnamefont{Ma}},
  \bibinfo{author}{\bibfnamefont{T.~A.} \bibnamefont{Cochran}},
  \bibinfo{author}{\bibfnamefont{X.}~\bibnamefont{Xu}},
  \bibinfo{author}{\bibfnamefont{S.~S.} \bibnamefont{Zhang}},
  \bibinfo{author}{\bibfnamefont{H.-J.} \bibnamefont{Tien}},
  \bibinfo{author}{\bibfnamefont{N.}~\bibnamefont{Shumiya}},
  \bibinfo{author}{\bibfnamefont{G.}~\bibnamefont{Cheng}},
  \bibinfo{author}{\bibfnamefont{K.}~\bibnamefont{Jiang}},
  \bibinfo{author}{\bibfnamefont{B.}~\bibnamefont{Lian}}, \bibnamefont{et~al.},
  \bibinfo{journal}{Nature} \textbf{\bibinfo{volume}{583}},
  \bibinfo{pages}{533} (\bibinfo{year}{2020}).

\bibitem[{\citenamefont{Liu et~al.}(2018)\citenamefont{Liu, Sun, Kumar,
  Muechler, Sun, Jiao, Yang, Liu, Liang, Xu et~al.}}]{Co3Sn2S2}
\bibinfo{author}{\bibfnamefont{E.}~\bibnamefont{Liu}},
  \bibinfo{author}{\bibfnamefont{Y.}~\bibnamefont{Sun}},
  \bibinfo{author}{\bibfnamefont{N.}~\bibnamefont{Kumar}},
  \bibinfo{author}{\bibfnamefont{L.}~\bibnamefont{Muechler}},
  \bibinfo{author}{\bibfnamefont{A.}~\bibnamefont{Sun}},
  \bibinfo{author}{\bibfnamefont{L.}~\bibnamefont{Jiao}},
  \bibinfo{author}{\bibfnamefont{S.-Y.} \bibnamefont{Yang}},
  \bibinfo{author}{\bibfnamefont{D.}~\bibnamefont{Liu}},
  \bibinfo{author}{\bibfnamefont{A.}~\bibnamefont{Liang}},
  \bibinfo{author}{\bibfnamefont{Q.}~\bibnamefont{Xu}}, \bibnamefont{et~al.},
  \bibinfo{journal}{Nat. Phys.} \textbf{\bibinfo{volume}{14}},
  \bibinfo{pages}{1125} (\bibinfo{year}{2018}).

\bibitem[{\citenamefont{Wang et~al.}(2018)\citenamefont{Wang, Xu, Lou, Liu, Li,
  Huang, Shen, Weng, Wang, and Lei}}]{Co3Sn2S2Lei}
\bibinfo{author}{\bibfnamefont{Q.}~\bibnamefont{Wang}},
  \bibinfo{author}{\bibfnamefont{Y.}~\bibnamefont{Xu}},
  \bibinfo{author}{\bibfnamefont{R.}~\bibnamefont{Lou}},
  \bibinfo{author}{\bibfnamefont{Z.}~\bibnamefont{Liu}},
  \bibinfo{author}{\bibfnamefont{M.}~\bibnamefont{Li}},
  \bibinfo{author}{\bibfnamefont{Y.}~\bibnamefont{Huang}},
  \bibinfo{author}{\bibfnamefont{D.}~\bibnamefont{Shen}},
  \bibinfo{author}{\bibfnamefont{H.}~\bibnamefont{Weng}},
  \bibinfo{author}{\bibfnamefont{S.}~\bibnamefont{Wang}}, \bibnamefont{and}
  \bibinfo{author}{\bibfnamefont{H.}~\bibnamefont{Lei}}, \bibinfo{journal}{Nat.
  Commun.} \textbf{\bibinfo{volume}{9}}, \bibinfo{pages}{1}
  (\bibinfo{year}{2018}).

\bibitem[{\citenamefont{Guin et~al.}(2019)\citenamefont{Guin, Vir, Zhang,
  Kumar, Watzman, Fu, Liu, Manna, Schnelle, Gooth et~al.}}]{CSSANE}
\bibinfo{author}{\bibfnamefont{S.~N.} \bibnamefont{Guin}},
  \bibinfo{author}{\bibfnamefont{P.}~\bibnamefont{Vir}},
  \bibinfo{author}{\bibfnamefont{Y.}~\bibnamefont{Zhang}},
  \bibinfo{author}{\bibfnamefont{N.}~\bibnamefont{Kumar}},
  \bibinfo{author}{\bibfnamefont{S.~J.} \bibnamefont{Watzman}},
  \bibinfo{author}{\bibfnamefont{C.}~\bibnamefont{Fu}},
  \bibinfo{author}{\bibfnamefont{E.}~\bibnamefont{Liu}},
  \bibinfo{author}{\bibfnamefont{K.}~\bibnamefont{Manna}},
  \bibinfo{author}{\bibfnamefont{W.}~\bibnamefont{Schnelle}},
  \bibinfo{author}{\bibfnamefont{J.}~\bibnamefont{Gooth}},
  \bibnamefont{et~al.}, \bibinfo{journal}{Adv. Mater.}
  \textbf{\bibinfo{volume}{31}}, \bibinfo{pages}{1806622}
  (\bibinfo{year}{2019}).

\bibitem[{\citenamefont{Nakatsuji et~al.}(2015)\citenamefont{Nakatsuji,
  Kiyohara, and Higo}}]{Mn3Sn}
\bibinfo{author}{\bibfnamefont{S.}~\bibnamefont{Nakatsuji}},
  \bibinfo{author}{\bibfnamefont{N.}~\bibnamefont{Kiyohara}}, \bibnamefont{and}
  \bibinfo{author}{\bibfnamefont{T.}~\bibnamefont{Higo}},
  \bibinfo{journal}{Nature} \textbf{\bibinfo{volume}{527}},
  \bibinfo{pages}{212} (\bibinfo{year}{2015}).

\bibitem[{\citenamefont{Ikhlas et~al.}(2017)\citenamefont{Ikhlas, Tomita,
  Koretsune, Suzuki, Nishio-Hamane, Arita, Otani, and Nakatsuji}}]{Mn3SnANE}
\bibinfo{author}{\bibfnamefont{M.}~\bibnamefont{Ikhlas}},
  \bibinfo{author}{\bibfnamefont{T.}~\bibnamefont{Tomita}},
  \bibinfo{author}{\bibfnamefont{T.}~\bibnamefont{Koretsune}},
  \bibinfo{author}{\bibfnamefont{M.-T.} \bibnamefont{Suzuki}},
  \bibinfo{author}{\bibfnamefont{D.}~\bibnamefont{Nishio-Hamane}},
  \bibinfo{author}{\bibfnamefont{R.}~\bibnamefont{Arita}},
  \bibinfo{author}{\bibfnamefont{Y.}~\bibnamefont{Otani}}, \bibnamefont{and}
  \bibinfo{author}{\bibfnamefont{S.}~\bibnamefont{Nakatsuji}},
  \bibinfo{journal}{Nat. Phys.} \textbf{\bibinfo{volume}{13}},
  \bibinfo{pages}{1085} (\bibinfo{year}{2017}).

\bibitem[{\citenamefont{Nayak et~al.}(2016)\citenamefont{Nayak, Fischer, Sun,
  Yan, Karel, Komarek, Shekhar, Kumar, Schnelle, K{\"u}bler et~al.}}]{Mn3Ge}
\bibinfo{author}{\bibfnamefont{A.~K.} \bibnamefont{Nayak}},
  \bibinfo{author}{\bibfnamefont{J.~E.} \bibnamefont{Fischer}},
  \bibinfo{author}{\bibfnamefont{Y.}~\bibnamefont{Sun}},
  \bibinfo{author}{\bibfnamefont{B.}~\bibnamefont{Yan}},
  \bibinfo{author}{\bibfnamefont{J.}~\bibnamefont{Karel}},
  \bibinfo{author}{\bibfnamefont{A.~C.} \bibnamefont{Komarek}},
  \bibinfo{author}{\bibfnamefont{C.}~\bibnamefont{Shekhar}},
  \bibinfo{author}{\bibfnamefont{N.}~\bibnamefont{Kumar}},
  \bibinfo{author}{\bibfnamefont{W.}~\bibnamefont{Schnelle}},
  \bibinfo{author}{\bibfnamefont{J.}~\bibnamefont{K{\"u}bler}},
  \bibnamefont{et~al.}, \bibinfo{journal}{Sci. Adv.}
  \textbf{\bibinfo{volume}{2}}, \bibinfo{pages}{e1501870}
  (\bibinfo{year}{2016}).

\bibitem[{\citenamefont{Ortiz et~al.}(2019)\citenamefont{Ortiz, Gomes, Morey,
  Winiarski, Bordelon, Mangum, Oswald, Rodriguez-Rivera, Neilson, Wilson
  et~al.}}]{AV3Sb5}
\bibinfo{author}{\bibfnamefont{B.~R.} \bibnamefont{Ortiz}},
  \bibinfo{author}{\bibfnamefont{L.~C.} \bibnamefont{Gomes}},
  \bibinfo{author}{\bibfnamefont{J.~R.} \bibnamefont{Morey}},
  \bibinfo{author}{\bibfnamefont{M.}~\bibnamefont{Winiarski}},
  \bibinfo{author}{\bibfnamefont{M.}~\bibnamefont{Bordelon}},
  \bibinfo{author}{\bibfnamefont{J.~S.} \bibnamefont{Mangum}},
  \bibinfo{author}{\bibfnamefont{I.~W.} \bibnamefont{Oswald}},
  \bibinfo{author}{\bibfnamefont{J.~A.} \bibnamefont{Rodriguez-Rivera}},
  \bibinfo{author}{\bibfnamefont{J.~R.} \bibnamefont{Neilson}},
  \bibinfo{author}{\bibfnamefont{S.~D.} \bibnamefont{Wilson}},
  \bibnamefont{et~al.}, \bibinfo{journal}{Phys. Rev. Mater.}
  \textbf{\bibinfo{volume}{3}}, \bibinfo{pages}{094407} (\bibinfo{year}{2019}).

\bibitem[{\citenamefont{Ortiz et~al.}(2021{\natexlab{a}})\citenamefont{Ortiz,
  Sarte, Kenney, Graf, Teicher, Seshadri, and Wilson}}]{KVSsupc}
\bibinfo{author}{\bibfnamefont{B.~R.} \bibnamefont{Ortiz}},
  \bibinfo{author}{\bibfnamefont{P.~M.} \bibnamefont{Sarte}},
  \bibinfo{author}{\bibfnamefont{E.~M.} \bibnamefont{Kenney}},
  \bibinfo{author}{\bibfnamefont{M.~J.} \bibnamefont{Graf}},
  \bibinfo{author}{\bibfnamefont{S.~M.} \bibnamefont{Teicher}},
  \bibinfo{author}{\bibfnamefont{R.}~\bibnamefont{Seshadri}}, \bibnamefont{and}
  \bibinfo{author}{\bibfnamefont{S.~D.} \bibnamefont{Wilson}},
  \bibinfo{journal}{Phys. Rev. Mater.} \textbf{\bibinfo{volume}{5}},
  \bibinfo{pages}{034801} (\bibinfo{year}{2021}{\natexlab{a}}).

\bibitem[{\citenamefont{Yin et~al.}(2021)\citenamefont{Yin, Tu, Gong, Fu, Yan,
  and Lei}}]{RVSsupc}
\bibinfo{author}{\bibfnamefont{Q.}~\bibnamefont{Yin}},
  \bibinfo{author}{\bibfnamefont{Z.}~\bibnamefont{Tu}},
  \bibinfo{author}{\bibfnamefont{C.}~\bibnamefont{Gong}},
  \bibinfo{author}{\bibfnamefont{Y.}~\bibnamefont{Fu}},
  \bibinfo{author}{\bibfnamefont{S.}~\bibnamefont{Yan}}, \bibnamefont{and}
  \bibinfo{author}{\bibfnamefont{H.}~\bibnamefont{Lei}},
  \bibinfo{journal}{Chin. Phys. Lett.} \textbf{\bibinfo{volume}{38}},
  \bibinfo{pages}{037403} (\bibinfo{year}{2021}).

\bibitem[{\citenamefont{Ortiz et~al.}(2020)\citenamefont{Ortiz, Teicher, Hu,
  Zuo, Sarte, Schueller, Abeykoon, Krogstad, Rosenkranz, Osborn
  et~al.}}]{CVSsupc}
\bibinfo{author}{\bibfnamefont{B.~R.} \bibnamefont{Ortiz}},
  \bibinfo{author}{\bibfnamefont{S.~M.} \bibnamefont{Teicher}},
  \bibinfo{author}{\bibfnamefont{Y.}~\bibnamefont{Hu}},
  \bibinfo{author}{\bibfnamefont{J.~L.} \bibnamefont{Zuo}},
  \bibinfo{author}{\bibfnamefont{P.~M.} \bibnamefont{Sarte}},
  \bibinfo{author}{\bibfnamefont{E.~C.} \bibnamefont{Schueller}},
  \bibinfo{author}{\bibfnamefont{A.~M.} \bibnamefont{Abeykoon}},
  \bibinfo{author}{\bibfnamefont{M.~J.} \bibnamefont{Krogstad}},
  \bibinfo{author}{\bibfnamefont{S.}~\bibnamefont{Rosenkranz}},
  \bibinfo{author}{\bibfnamefont{R.}~\bibnamefont{Osborn}},
  \bibnamefont{et~al.}, \bibinfo{journal}{Phys. Rev. Lett.}
  \textbf{\bibinfo{volume}{125}}, \bibinfo{pages}{247002}
  (\bibinfo{year}{2020}).

\bibitem[{\citenamefont{Tan et~al.}(2021)\citenamefont{Tan, Liu, Wang, and
  Yan}}]{CVSDFTYan}
\bibinfo{author}{\bibfnamefont{H.}~\bibnamefont{Tan}},
  \bibinfo{author}{\bibfnamefont{Y.}~\bibnamefont{Liu}},
  \bibinfo{author}{\bibfnamefont{Z.}~\bibnamefont{Wang}}, \bibnamefont{and}
  \bibinfo{author}{\bibfnamefont{B.}~\bibnamefont{Yan}},
  \bibinfo{journal}{Phys. Rev. Lett.} \textbf{\bibinfo{volume}{127}},
  \bibinfo{pages}{046401} (\bibinfo{year}{2021}).

\bibitem[{\citenamefont{Jiang et~al.}(2021)\citenamefont{Jiang, Yin, Denner,
  Shumiya, Ortiz, Xu, Guguchia, He, Hossain, Liu et~al.}}]{KVSCDW}
\bibinfo{author}{\bibfnamefont{Y.-X.} \bibnamefont{Jiang}},
  \bibinfo{author}{\bibfnamefont{J.-X.} \bibnamefont{Yin}},
  \bibinfo{author}{\bibfnamefont{M.~M.} \bibnamefont{Denner}},
  \bibinfo{author}{\bibfnamefont{N.}~\bibnamefont{Shumiya}},
  \bibinfo{author}{\bibfnamefont{B.~R.} \bibnamefont{Ortiz}},
  \bibinfo{author}{\bibfnamefont{G.}~\bibnamefont{Xu}},
  \bibinfo{author}{\bibfnamefont{Z.}~\bibnamefont{Guguchia}},
  \bibinfo{author}{\bibfnamefont{J.}~\bibnamefont{He}},
  \bibinfo{author}{\bibfnamefont{M.~S.} \bibnamefont{Hossain}},
  \bibinfo{author}{\bibfnamefont{X.}~\bibnamefont{Liu}}, \bibnamefont{et~al.},
  \bibinfo{journal}{Nat. Mater.} pp. \bibinfo{pages}{1--5}
  (\bibinfo{year}{2021}).

\bibitem[{\citenamefont{Chen et~al.}(2021)\citenamefont{Chen, Yang, Hu, Zhao,
  Yuan, Xing, Qian, Huang, Li, Ye et~al.}}]{CVSSTM1}
\bibinfo{author}{\bibfnamefont{H.}~\bibnamefont{Chen}},
  \bibinfo{author}{\bibfnamefont{H.}~\bibnamefont{Yang}},
  \bibinfo{author}{\bibfnamefont{B.}~\bibnamefont{Hu}},
  \bibinfo{author}{\bibfnamefont{Z.}~\bibnamefont{Zhao}},
  \bibinfo{author}{\bibfnamefont{J.}~\bibnamefont{Yuan}},
  \bibinfo{author}{\bibfnamefont{Y.}~\bibnamefont{Xing}},
  \bibinfo{author}{\bibfnamefont{G.}~\bibnamefont{Qian}},
  \bibinfo{author}{\bibfnamefont{Z.}~\bibnamefont{Huang}},
  \bibinfo{author}{\bibfnamefont{G.}~\bibnamefont{Li}},
  \bibinfo{author}{\bibfnamefont{Y.}~\bibnamefont{Ye}}, \bibnamefont{et~al.},
  \bibinfo{journal}{Nature} pp. \bibinfo{pages}{1--9} (\bibinfo{year}{2021}).

\bibitem[{\citenamefont{Zhao et~al.}(2021)\citenamefont{Zhao, Li, Ortiz,
  Teicher, Park, Ye, Wang, Balents, Wilson, and Zeljkovic}}]{CVSSTM2}
\bibinfo{author}{\bibfnamefont{H.}~\bibnamefont{Zhao}},
  \bibinfo{author}{\bibfnamefont{H.}~\bibnamefont{Li}},
  \bibinfo{author}{\bibfnamefont{B.~R.} \bibnamefont{Ortiz}},
  \bibinfo{author}{\bibfnamefont{S.~M.} \bibnamefont{Teicher}},
  \bibinfo{author}{\bibfnamefont{T.}~\bibnamefont{Park}},
  \bibinfo{author}{\bibfnamefont{M.}~\bibnamefont{Ye}},
  \bibinfo{author}{\bibfnamefont{Z.}~\bibnamefont{Wang}},
  \bibinfo{author}{\bibfnamefont{L.}~\bibnamefont{Balents}},
  \bibinfo{author}{\bibfnamefont{S.~D.} \bibnamefont{Wilson}},
  \bibnamefont{and}
  \bibinfo{author}{\bibfnamefont{I.}~\bibnamefont{Zeljkovic}},
  \bibinfo{journal}{arXiv preprint arXiv:2103.03118}  (\bibinfo{year}{2021}).

\bibitem[{\citenamefont{Liang et~al.}(2021)\citenamefont{Liang, Hou, Zhang, Ma,
  Wu, Zhang, Yu, Ying, Jiang, Shan et~al.}}]{CVSSTMXHChen}
\bibinfo{author}{\bibfnamefont{Z.}~\bibnamefont{Liang}},
  \bibinfo{author}{\bibfnamefont{X.}~\bibnamefont{Hou}},
  \bibinfo{author}{\bibfnamefont{F.}~\bibnamefont{Zhang}},
  \bibinfo{author}{\bibfnamefont{W.}~\bibnamefont{Ma}},
  \bibinfo{author}{\bibfnamefont{P.}~\bibnamefont{Wu}},
  \bibinfo{author}{\bibfnamefont{Z.}~\bibnamefont{Zhang}},
  \bibinfo{author}{\bibfnamefont{F.}~\bibnamefont{Yu}},
  \bibinfo{author}{\bibfnamefont{J.-J.} \bibnamefont{Ying}},
  \bibinfo{author}{\bibfnamefont{K.}~\bibnamefont{Jiang}},
  \bibinfo{author}{\bibfnamefont{L.}~\bibnamefont{Shan}}, \bibnamefont{et~al.},
  \bibinfo{journal}{Phys. Rev. X} \textbf{\bibinfo{volume}{11}},
  \bibinfo{pages}{031026} (\bibinfo{year}{2021}).

\bibitem[{\citenamefont{Yang et~al.}(2020)\citenamefont{Yang, Wang, Ortiz, Liu,
  Gayles, Derunova, Gonzalez-Hernandez, {\v{S}}mejkal, Chen, Parkin
  et~al.}}]{KVSAHE}
\bibinfo{author}{\bibfnamefont{S.}~\bibnamefont{Yang}},
  \bibinfo{author}{\bibfnamefont{Y.}~\bibnamefont{Wang}},
  \bibinfo{author}{\bibfnamefont{B.}~\bibnamefont{Ortiz}},
  \bibinfo{author}{\bibfnamefont{D.}~\bibnamefont{Liu}},
  \bibinfo{author}{\bibfnamefont{J.}~\bibnamefont{Gayles}},
  \bibinfo{author}{\bibfnamefont{E.}~\bibnamefont{Derunova}},
  \bibinfo{author}{\bibfnamefont{R.}~\bibnamefont{Gonzalez-Hernandez}},
  \bibinfo{author}{\bibfnamefont{L.}~\bibnamefont{{\v{S}}mejkal}},
  \bibinfo{author}{\bibfnamefont{Y.}~\bibnamefont{Chen}},
  \bibinfo{author}{\bibfnamefont{S.}~\bibnamefont{Parkin}},
  \bibnamefont{et~al.}, \bibinfo{journal}{Sci. Adv.}
  \textbf{\bibinfo{volume}{6}}, \bibinfo{pages}{eabb6003}
  (\bibinfo{year}{2020}).

\bibitem[{\citenamefont{Yu et~al.}(2021{\natexlab{a}})\citenamefont{Yu, Wu,
  Wang, Lei, Zhuo, Ying, and Chen}}]{CVSAHE}
\bibinfo{author}{\bibfnamefont{F.~H.} \bibnamefont{Yu}},
  \bibinfo{author}{\bibfnamefont{T.}~\bibnamefont{Wu}},
  \bibinfo{author}{\bibfnamefont{Z.~Y.} \bibnamefont{Wang}},
  \bibinfo{author}{\bibfnamefont{B.}~\bibnamefont{Lei}},
  \bibinfo{author}{\bibfnamefont{W.~Z.} \bibnamefont{Zhuo}},
  \bibinfo{author}{\bibfnamefont{J.~J.} \bibnamefont{Ying}}, \bibnamefont{and}
  \bibinfo{author}{\bibfnamefont{X.~H.} \bibnamefont{Chen}},
  \bibinfo{journal}{Phys. Rev. B} \textbf{\bibinfo{volume}{104}},
  \bibinfo{pages}{L041103} (\bibinfo{year}{2021}{\natexlab{a}}).

\bibitem[{\citenamefont{Kenney et~al.}(2021)\citenamefont{Kenney, Ortiz, Wang,
  Wilson, and Graf}}]{CVSuSR1}
\bibinfo{author}{\bibfnamefont{E.~M.} \bibnamefont{Kenney}},
  \bibinfo{author}{\bibfnamefont{B.~R.} \bibnamefont{Ortiz}},
  \bibinfo{author}{\bibfnamefont{C.}~\bibnamefont{Wang}},
  \bibinfo{author}{\bibfnamefont{S.~D.} \bibnamefont{Wilson}},
  \bibnamefont{and} \bibinfo{author}{\bibfnamefont{M.~J.} \bibnamefont{Graf}},
  \bibinfo{journal}{J. Phys. Condens. Matter} \textbf{\bibinfo{volume}{33}},
  \bibinfo{pages}{235801} (\bibinfo{year}{2021}).

\bibitem[{\citenamefont{Mielke~III et~al.}(2021)\citenamefont{Mielke~III, Das,
  Yin, Liu, Gupta, Wang, Jiang, Medarde, Wu, Lei et~al.}}]{KVSuSR}
\bibinfo{author}{\bibfnamefont{C.}~\bibnamefont{Mielke~III}},
  \bibinfo{author}{\bibfnamefont{D.}~\bibnamefont{Das}},
  \bibinfo{author}{\bibfnamefont{J.-X.} \bibnamefont{Yin}},
  \bibinfo{author}{\bibfnamefont{H.}~\bibnamefont{Liu}},
  \bibinfo{author}{\bibfnamefont{R.}~\bibnamefont{Gupta}},
  \bibinfo{author}{\bibfnamefont{C.}~\bibnamefont{Wang}},
  \bibinfo{author}{\bibfnamefont{Y.-X.} \bibnamefont{Jiang}},
  \bibinfo{author}{\bibfnamefont{M.}~\bibnamefont{Medarde}},
  \bibinfo{author}{\bibfnamefont{X.}~\bibnamefont{Wu}},
  \bibinfo{author}{\bibfnamefont{H.}~\bibnamefont{Lei}}, \bibnamefont{et~al.},
  \bibinfo{journal}{arXiv preprint arXiv:2106.13443}  (\bibinfo{year}{2021}).

\bibitem[{\citenamefont{Yu et~al.}(2021{\natexlab{b}})\citenamefont{Yu, Wang,
  Zhang, Sander, Ni, Lu, Ma, Wang, Zhao, Chen et~al.}}]{CVSuSR2}
\bibinfo{author}{\bibfnamefont{L.}~\bibnamefont{Yu}},
  \bibinfo{author}{\bibfnamefont{C.}~\bibnamefont{Wang}},
  \bibinfo{author}{\bibfnamefont{Y.}~\bibnamefont{Zhang}},
  \bibinfo{author}{\bibfnamefont{M.}~\bibnamefont{Sander}},
  \bibinfo{author}{\bibfnamefont{S.}~\bibnamefont{Ni}},
  \bibinfo{author}{\bibfnamefont{Z.}~\bibnamefont{Lu}},
  \bibinfo{author}{\bibfnamefont{S.}~\bibnamefont{Ma}},
  \bibinfo{author}{\bibfnamefont{Z.}~\bibnamefont{Wang}},
  \bibinfo{author}{\bibfnamefont{Z.}~\bibnamefont{Zhao}},
  \bibinfo{author}{\bibfnamefont{H.}~\bibnamefont{Chen}}, \bibnamefont{et~al.},
  \bibinfo{journal}{arXiv preprint arXiv:2107.10714}
  (\bibinfo{year}{2021}{\natexlab{b}}).

\bibitem[{\citenamefont{Hosur and Qi}(2013)}]{WeylQi}
\bibinfo{author}{\bibfnamefont{P.}~\bibnamefont{Hosur}} \bibnamefont{and}
  \bibinfo{author}{\bibfnamefont{X.}~\bibnamefont{Qi}}, \bibinfo{journal}{C.R.
  Phys.} \textbf{\bibinfo{volume}{14}}, \bibinfo{pages}{857}
  (\bibinfo{year}{2013}).

\bibitem[{\citenamefont{Xiao et~al.}(2010)\citenamefont{Xiao, Chang, and
  Niu}}]{BerryRMP}
\bibinfo{author}{\bibfnamefont{D.}~\bibnamefont{Xiao}},
  \bibinfo{author}{\bibfnamefont{M.-C.} \bibnamefont{Chang}}, \bibnamefont{and}
  \bibinfo{author}{\bibfnamefont{Q.}~\bibnamefont{Niu}}, \bibinfo{journal}{Rev.
  Mod. Phys.} \textbf{\bibinfo{volume}{82}}, \bibinfo{pages}{1959}
  (\bibinfo{year}{2010}).

\bibitem[{\citenamefont{Ziman}(1960)}]{zimanE&P}
\bibinfo{author}{\bibfnamefont{J.~M.} \bibnamefont{Ziman}},
  \emph{\bibinfo{title}{Electrons and phonons}} (\bibinfo{publisher}{Clarendon
  Press, Oxford}, \bibinfo{year}{1960}).

\bibitem[{\citenamefont{Liang et~al.}(2017)\citenamefont{Liang, Lin, Gibson,
  Gao, Hirschberger, Liu, Cava, and Ong}}]{Cd3As2ANE}
\bibinfo{author}{\bibfnamefont{T.}~\bibnamefont{Liang}},
  \bibinfo{author}{\bibfnamefont{J.}~\bibnamefont{Lin}},
  \bibinfo{author}{\bibfnamefont{Q.}~\bibnamefont{Gibson}},
  \bibinfo{author}{\bibfnamefont{T.}~\bibnamefont{Gao}},
  \bibinfo{author}{\bibfnamefont{M.}~\bibnamefont{Hirschberger}},
  \bibinfo{author}{\bibfnamefont{M.}~\bibnamefont{Liu}},
  \bibinfo{author}{\bibfnamefont{R.~J.} \bibnamefont{Cava}}, \bibnamefont{and}
  \bibinfo{author}{\bibfnamefont{N.~P.} \bibnamefont{Ong}},
  \bibinfo{journal}{Phys. Rev. Lett.} \textbf{\bibinfo{volume}{118}},
  \bibinfo{pages}{136601} (\bibinfo{year}{2017}).

\bibitem[{\citenamefont{Zhang et~al.}(2019)\citenamefont{Zhang, Wang, Guo, Zhu,
  Zhang, Yang, Wang, Qu, Pi, Lu et~al.}}]{ZrTe5ANE}
\bibinfo{author}{\bibfnamefont{J.}~\bibnamefont{Zhang}},
  \bibinfo{author}{\bibfnamefont{C.}~\bibnamefont{Wang}},
  \bibinfo{author}{\bibfnamefont{C.}~\bibnamefont{Guo}},
  \bibinfo{author}{\bibfnamefont{X.}~\bibnamefont{Zhu}},
  \bibinfo{author}{\bibfnamefont{Y.}~\bibnamefont{Zhang}},
  \bibinfo{author}{\bibfnamefont{J.}~\bibnamefont{Yang}},
  \bibinfo{author}{\bibfnamefont{Y.}~\bibnamefont{Wang}},
  \bibinfo{author}{\bibfnamefont{Z.}~\bibnamefont{Qu}},
  \bibinfo{author}{\bibfnamefont{L.}~\bibnamefont{Pi}},
  \bibinfo{author}{\bibfnamefont{H.-Z.} \bibnamefont{Lu}},
  \bibnamefont{et~al.}, \bibinfo{journal}{Phys. Rev. Lett.}
  \textbf{\bibinfo{volume}{123}}, \bibinfo{pages}{196602}
  (\bibinfo{year}{2019}).

\bibitem[{\citenamefont{Zhu et~al.}(2015)\citenamefont{Zhu, Lin, Liu,
  Fauqu{\'e}, Tao, Yang, Shi, and Behnia}}]{WTeBehnia}
\bibinfo{author}{\bibfnamefont{Z.}~\bibnamefont{Zhu}},
  \bibinfo{author}{\bibfnamefont{X.}~\bibnamefont{Lin}},
  \bibinfo{author}{\bibfnamefont{J.}~\bibnamefont{Liu}},
  \bibinfo{author}{\bibfnamefont{B.}~\bibnamefont{Fauqu{\'e}}},
  \bibinfo{author}{\bibfnamefont{Q.}~\bibnamefont{Tao}},
  \bibinfo{author}{\bibfnamefont{C.}~\bibnamefont{Yang}},
  \bibinfo{author}{\bibfnamefont{Y.}~\bibnamefont{Shi}}, \bibnamefont{and}
  \bibinfo{author}{\bibfnamefont{K.}~\bibnamefont{Behnia}},
  \bibinfo{journal}{Phys. Rev. Lett.} \textbf{\bibinfo{volume}{114}},
  \bibinfo{pages}{176601} (\bibinfo{year}{2015}).

\bibitem[{\citenamefont{Liang et~al.}(2013)\citenamefont{Liang, Gibson, Xiong,
  Hirschberger, Koduvayur, Cava, and Ong}}]{PbSe}
\bibinfo{author}{\bibfnamefont{T.}~\bibnamefont{Liang}},
  \bibinfo{author}{\bibfnamefont{Q.}~\bibnamefont{Gibson}},
  \bibinfo{author}{\bibfnamefont{J.}~\bibnamefont{Xiong}},
  \bibinfo{author}{\bibfnamefont{M.}~\bibnamefont{Hirschberger}},
  \bibinfo{author}{\bibfnamefont{S.~P.} \bibnamefont{Koduvayur}},
  \bibinfo{author}{\bibfnamefont{R.~J.} \bibnamefont{Cava}}, \bibnamefont{and}
  \bibinfo{author}{\bibfnamefont{N.~P.} \bibnamefont{Ong}},
  \bibinfo{journal}{Nat. Commun.} \textbf{\bibinfo{volume}{4}},
  \bibinfo{pages}{1} (\bibinfo{year}{2013}).

\bibitem[{\citenamefont{Watzman et~al.}(2018)\citenamefont{Watzman, McCormick,
  Shekhar, Wu, Sun, Prakash, Felser, Trivedi, and Heremans}}]{NbPNernst}
\bibinfo{author}{\bibfnamefont{S.~J.} \bibnamefont{Watzman}},
  \bibinfo{author}{\bibfnamefont{T.~M.} \bibnamefont{McCormick}},
  \bibinfo{author}{\bibfnamefont{C.}~\bibnamefont{Shekhar}},
  \bibinfo{author}{\bibfnamefont{S.-C.} \bibnamefont{Wu}},
  \bibinfo{author}{\bibfnamefont{Y.}~\bibnamefont{Sun}},
  \bibinfo{author}{\bibfnamefont{A.}~\bibnamefont{Prakash}},
  \bibinfo{author}{\bibfnamefont{C.}~\bibnamefont{Felser}},
  \bibinfo{author}{\bibfnamefont{N.}~\bibnamefont{Trivedi}}, \bibnamefont{and}
  \bibinfo{author}{\bibfnamefont{J.~P.} \bibnamefont{Heremans}},
  \bibinfo{journal}{Phys. Rev. B} \textbf{\bibinfo{volume}{97}},
  \bibinfo{pages}{161404} (\bibinfo{year}{2018}).

\bibitem[{\citenamefont{Gan et~al.}(2021)\citenamefont{Gan, Xia, Zhang, Yang,
  Mi, Wang, Chai, Guo, Zhou, and He}}]{CVSTE}
\bibinfo{author}{\bibfnamefont{Y.}~\bibnamefont{Gan}},
  \bibinfo{author}{\bibfnamefont{W.}~\bibnamefont{Xia}},
  \bibinfo{author}{\bibfnamefont{L.}~\bibnamefont{Zhang}},
  \bibinfo{author}{\bibfnamefont{K.}~\bibnamefont{Yang}},
  \bibinfo{author}{\bibfnamefont{X.}~\bibnamefont{Mi}},
  \bibinfo{author}{\bibfnamefont{A.}~\bibnamefont{Wang}},
  \bibinfo{author}{\bibfnamefont{Y.}~\bibnamefont{Chai}},
  \bibinfo{author}{\bibfnamefont{Y.}~\bibnamefont{Guo}},
  \bibinfo{author}{\bibfnamefont{X.}~\bibnamefont{Zhou}}, \bibnamefont{and}
  \bibinfo{author}{\bibfnamefont{M.}~\bibnamefont{He}}, \bibinfo{journal}{arXiv
  preprint arXiv:2110.00289}  (\bibinfo{year}{2021}).

\bibitem[{\citenamefont{Ortiz et~al.}(2021{\natexlab{b}})\citenamefont{Ortiz,
  Teicher, Kautzsch, Sarte, Ruff, Seshadri, and Wilson}}]{CVSFSmapping}
\bibinfo{author}{\bibfnamefont{B.~R.} \bibnamefont{Ortiz}},
  \bibinfo{author}{\bibfnamefont{S.~M.} \bibnamefont{Teicher}},
  \bibinfo{author}{\bibfnamefont{L.}~\bibnamefont{Kautzsch}},
  \bibinfo{author}{\bibfnamefont{P.~M.} \bibnamefont{Sarte}},
  \bibinfo{author}{\bibfnamefont{J.~P.} \bibnamefont{Ruff}},
  \bibinfo{author}{\bibfnamefont{R.}~\bibnamefont{Seshadri}}, \bibnamefont{and}
  \bibinfo{author}{\bibfnamefont{S.~D.} \bibnamefont{Wilson}},
  \bibinfo{journal}{arXiv preprint arXiv:2104.07230}
  (\bibinfo{year}{2021}{\natexlab{b}}).

\bibitem[{\citenamefont{Shoenberg}(1984)}]{shoenberg}
\bibinfo{author}{\bibfnamefont{D.}~\bibnamefont{Shoenberg}},
  \emph{\bibinfo{title}{Magnetic oscillations in metals}}
  (\bibinfo{publisher}{Cambridge university press}, \bibinfo{year}{1984}).

\bibitem[{\citenamefont{Fauqu{\'e} et~al.}(2013)\citenamefont{Fauqu{\'e},
  Butch, Syers, Paglione, Wiedmann, Collaudin, Grena, Zeitler, and
  Behnia}}]{Bi2Se3Behnia}
\bibinfo{author}{\bibfnamefont{B.}~\bibnamefont{Fauqu{\'e}}},
  \bibinfo{author}{\bibfnamefont{N.~P.} \bibnamefont{Butch}},
  \bibinfo{author}{\bibfnamefont{P.}~\bibnamefont{Syers}},
  \bibinfo{author}{\bibfnamefont{J.}~\bibnamefont{Paglione}},
  \bibinfo{author}{\bibfnamefont{S.}~\bibnamefont{Wiedmann}},
  \bibinfo{author}{\bibfnamefont{A.}~\bibnamefont{Collaudin}},
  \bibinfo{author}{\bibfnamefont{B.}~\bibnamefont{Grena}},
  \bibinfo{author}{\bibfnamefont{U.}~\bibnamefont{Zeitler}}, \bibnamefont{and}
  \bibinfo{author}{\bibfnamefont{K.}~\bibnamefont{Behnia}},
  \bibinfo{journal}{Phys. Rev. B} \textbf{\bibinfo{volume}{87}},
  \bibinfo{pages}{035133} (\bibinfo{year}{2013}).

\bibitem[{\citenamefont{Fu et~al.}(2021)\citenamefont{Fu, Zhao, Chen, Yin, Tu,
  Gong, Xi, Zhu, Sun, Liu et~al.}}]{CVSLei}
\bibinfo{author}{\bibfnamefont{Y.}~\bibnamefont{Fu}},
  \bibinfo{author}{\bibfnamefont{N.}~\bibnamefont{Zhao}},
  \bibinfo{author}{\bibfnamefont{Z.}~\bibnamefont{Chen}},
  \bibinfo{author}{\bibfnamefont{Q.}~\bibnamefont{Yin}},
  \bibinfo{author}{\bibfnamefont{Z.}~\bibnamefont{Tu}},
  \bibinfo{author}{\bibfnamefont{C.}~\bibnamefont{Gong}},
  \bibinfo{author}{\bibfnamefont{C.}~\bibnamefont{Xi}},
  \bibinfo{author}{\bibfnamefont{X.}~\bibnamefont{Zhu}},
  \bibinfo{author}{\bibfnamefont{Y.}~\bibnamefont{Sun}},
  \bibinfo{author}{\bibfnamefont{K.}~\bibnamefont{Liu}}, \bibnamefont{et~al.},
  \bibinfo{journal}{arXiv preprint arXiv:2104.08193}  (\bibinfo{year}{2021}).

\bibitem[{\citenamefont{Zhou et~al.}(2021)\citenamefont{Zhou, Li, Fan, Hao,
  Dai, Wang, Yao, and Wen}}]{CVSgapWen}
\bibinfo{author}{\bibfnamefont{X.}~\bibnamefont{Zhou}},
  \bibinfo{author}{\bibfnamefont{Y.}~\bibnamefont{Li}},
  \bibinfo{author}{\bibfnamefont{X.}~\bibnamefont{Fan}},
  \bibinfo{author}{\bibfnamefont{J.}~\bibnamefont{Hao}},
  \bibinfo{author}{\bibfnamefont{Y.}~\bibnamefont{Dai}},
  \bibinfo{author}{\bibfnamefont{Z.}~\bibnamefont{Wang}},
  \bibinfo{author}{\bibfnamefont{Y.}~\bibnamefont{Yao}}, \bibnamefont{and}
  \bibinfo{author}{\bibfnamefont{H.-H.} \bibnamefont{Wen}},
  \bibinfo{journal}{arXiv preprint arXiv:2104.01015}  (\bibinfo{year}{2021}).

\bibitem[{\citenamefont{Kang et~al.}(2021)\citenamefont{Kang, Fang, Kim, Ortiz,
  Yoo, Park, Wilson, Park, and Comin}}]{CVSgapARPES1}
\bibinfo{author}{\bibfnamefont{M.}~\bibnamefont{Kang}},
  \bibinfo{author}{\bibfnamefont{S.}~\bibnamefont{Fang}},
  \bibinfo{author}{\bibfnamefont{J.-K.} \bibnamefont{Kim}},
  \bibinfo{author}{\bibfnamefont{B.~R.} \bibnamefont{Ortiz}},
  \bibinfo{author}{\bibfnamefont{J.}~\bibnamefont{Yoo}},
  \bibinfo{author}{\bibfnamefont{B.-G.} \bibnamefont{Park}},
  \bibinfo{author}{\bibfnamefont{S.~D.} \bibnamefont{Wilson}},
  \bibinfo{author}{\bibfnamefont{J.-H.} \bibnamefont{Park}}, \bibnamefont{and}
  \bibinfo{author}{\bibfnamefont{R.}~\bibnamefont{Comin}},
  \bibinfo{journal}{arXiv preprint arXiv:2105.01689}  (\bibinfo{year}{2021}).

\bibitem[{\citenamefont{Nakayama et~al.}(2021)\citenamefont{Nakayama, Li, Liu,
  Wang, Takahashi, Yao, and Sato}}]{CVSgapARPES3}
\bibinfo{author}{\bibfnamefont{K.}~\bibnamefont{Nakayama}},
  \bibinfo{author}{\bibfnamefont{Y.}~\bibnamefont{Li}},
  \bibinfo{author}{\bibfnamefont{M.}~\bibnamefont{Liu}},
  \bibinfo{author}{\bibfnamefont{Z.}~\bibnamefont{Wang}},
  \bibinfo{author}{\bibfnamefont{T.}~\bibnamefont{Takahashi}},
  \bibinfo{author}{\bibfnamefont{Y.}~\bibnamefont{Yao}}, \bibnamefont{and}
  \bibinfo{author}{\bibfnamefont{T.}~\bibnamefont{Sato}},
  \bibinfo{journal}{arXiv preprint arXiv:2104.08042}  (\bibinfo{year}{2021}).

\bibitem[{\citenamefont{Wang et~al.}(2021{\natexlab{a}})\citenamefont{Wang, Ma,
  Zhang, Yang, Zhao, Ou, Zhu, Ni, Lu, Chen et~al.}}]{CVSgapARPES2}
\bibinfo{author}{\bibfnamefont{Z.}~\bibnamefont{Wang}},
  \bibinfo{author}{\bibfnamefont{S.}~\bibnamefont{Ma}},
  \bibinfo{author}{\bibfnamefont{Y.}~\bibnamefont{Zhang}},
  \bibinfo{author}{\bibfnamefont{H.}~\bibnamefont{Yang}},
  \bibinfo{author}{\bibfnamefont{Z.}~\bibnamefont{Zhao}},
  \bibinfo{author}{\bibfnamefont{Y.}~\bibnamefont{Ou}},
  \bibinfo{author}{\bibfnamefont{Y.}~\bibnamefont{Zhu}},
  \bibinfo{author}{\bibfnamefont{S.}~\bibnamefont{Ni}},
  \bibinfo{author}{\bibfnamefont{Z.}~\bibnamefont{Lu}},
  \bibinfo{author}{\bibfnamefont{H.}~\bibnamefont{Chen}}, \bibnamefont{et~al.},
  \bibinfo{journal}{arXiv preprint arXiv:2104.05556}
  (\bibinfo{year}{2021}{\natexlab{a}}).

\bibitem[{\citenamefont{Luo et~al.}(2021)\citenamefont{Luo, Gao, Liu, Gu, Wu,
  Yi, Jia, Wu, Luo, Xu et~al.}}]{KVSgapARPES}
\bibinfo{author}{\bibfnamefont{H.}~\bibnamefont{Luo}},
  \bibinfo{author}{\bibfnamefont{Q.}~\bibnamefont{Gao}},
  \bibinfo{author}{\bibfnamefont{H.}~\bibnamefont{Liu}},
  \bibinfo{author}{\bibfnamefont{Y.}~\bibnamefont{Gu}},
  \bibinfo{author}{\bibfnamefont{D.}~\bibnamefont{Wu}},
  \bibinfo{author}{\bibfnamefont{C.}~\bibnamefont{Yi}},
  \bibinfo{author}{\bibfnamefont{J.}~\bibnamefont{Jia}},
  \bibinfo{author}{\bibfnamefont{S.}~\bibnamefont{Wu}},
  \bibinfo{author}{\bibfnamefont{X.}~\bibnamefont{Luo}},
  \bibinfo{author}{\bibfnamefont{Y.}~\bibnamefont{Xu}}, \bibnamefont{et~al.},
  \bibinfo{journal}{arXiv preprint arXiv:2107.02688}  (\bibinfo{year}{2021}).

\bibitem[{\citenamefont{Liu et~al.}(2021)\citenamefont{Liu, Zhao, Yin, Gong,
  Tu, Li, Song, Liu, Shen, Huang et~al.}}]{RVSgapARPES}
\bibinfo{author}{\bibfnamefont{Z.}~\bibnamefont{Liu}},
  \bibinfo{author}{\bibfnamefont{N.}~\bibnamefont{Zhao}},
  \bibinfo{author}{\bibfnamefont{Q.}~\bibnamefont{Yin}},
  \bibinfo{author}{\bibfnamefont{C.}~\bibnamefont{Gong}},
  \bibinfo{author}{\bibfnamefont{Z.}~\bibnamefont{Tu}},
  \bibinfo{author}{\bibfnamefont{M.}~\bibnamefont{Li}},
  \bibinfo{author}{\bibfnamefont{W.}~\bibnamefont{Song}},
  \bibinfo{author}{\bibfnamefont{Z.}~\bibnamefont{Liu}},
  \bibinfo{author}{\bibfnamefont{D.}~\bibnamefont{Shen}},
  \bibinfo{author}{\bibfnamefont{Y.}~\bibnamefont{Huang}},
  \bibnamefont{et~al.}, \bibinfo{journal}{Phys. Rev. X}
  \textbf{\bibinfo{volume}{11}}, \bibinfo{pages}{041010}
  (\bibinfo{year}{2021}).

\bibitem[{\citenamefont{Wang et~al.}(2021{\natexlab{b}})\citenamefont{Wang,
  Jiang, Yin, Li, Wang, Huang, Shao, Liu, Zhu, Shumiya et~al.}}]{CVSCDW}
\bibinfo{author}{\bibfnamefont{Z.}~\bibnamefont{Wang}},
  \bibinfo{author}{\bibfnamefont{Y.-X.} \bibnamefont{Jiang}},
  \bibinfo{author}{\bibfnamefont{J.-X.} \bibnamefont{Yin}},
  \bibinfo{author}{\bibfnamefont{Y.}~\bibnamefont{Li}},
  \bibinfo{author}{\bibfnamefont{G.-Y.} \bibnamefont{Wang}},
  \bibinfo{author}{\bibfnamefont{H.-L.} \bibnamefont{Huang}},
  \bibinfo{author}{\bibfnamefont{S.}~\bibnamefont{Shao}},
  \bibinfo{author}{\bibfnamefont{J.}~\bibnamefont{Liu}},
  \bibinfo{author}{\bibfnamefont{P.}~\bibnamefont{Zhu}},
  \bibinfo{author}{\bibfnamefont{N.}~\bibnamefont{Shumiya}},
  \bibnamefont{et~al.}, \bibinfo{journal}{Phys. Rev. B}
  \textbf{\bibinfo{volume}{104}}, \bibinfo{pages}{075148}
  (\bibinfo{year}{2021}{\natexlab{b}}).

\bibitem[{\citenamefont{Feng et~al.}(2021)\citenamefont{Feng, Jiang, Wang, and
  Hu}}]{AVSCDW}
\bibinfo{author}{\bibfnamefont{X.}~\bibnamefont{Feng}},
  \bibinfo{author}{\bibfnamefont{K.}~\bibnamefont{Jiang}},
  \bibinfo{author}{\bibfnamefont{Z.}~\bibnamefont{Wang}}, \bibnamefont{and}
  \bibinfo{author}{\bibfnamefont{J.}~\bibnamefont{Hu}}, \bibinfo{journal}{Sci.
  Bull.}  (\bibinfo{year}{2021}).

\bibitem[{\citenamefont{Denner et~al.}(2021)\citenamefont{Denner, Thomale, and
  Neupert}}]{AVSCDW2}
\bibinfo{author}{\bibfnamefont{M.~M.} \bibnamefont{Denner}},
  \bibinfo{author}{\bibfnamefont{R.}~\bibnamefont{Thomale}}, \bibnamefont{and}
  \bibinfo{author}{\bibfnamefont{T.}~\bibnamefont{Neupert}},
  \bibinfo{journal}{arXiv preprint arXiv:2103.14045}  (\bibinfo{year}{2021}).

\bibitem[{\citenamefont{Zheng et~al.}(2021)\citenamefont{Zheng, Chen, Tan,
  Wang, Zhu, Albarakati, Algarni, Partridge, Farrar, Zhou et~al.}}]{CVSAHEgate}
\bibinfo{author}{\bibfnamefont{G.}~\bibnamefont{Zheng}},
  \bibinfo{author}{\bibfnamefont{Z.}~\bibnamefont{Chen}},
  \bibinfo{author}{\bibfnamefont{C.}~\bibnamefont{Tan}},
  \bibinfo{author}{\bibfnamefont{M.}~\bibnamefont{Wang}},
  \bibinfo{author}{\bibfnamefont{X.}~\bibnamefont{Zhu}},
  \bibinfo{author}{\bibfnamefont{S.}~\bibnamefont{Albarakati}},
  \bibinfo{author}{\bibfnamefont{M.}~\bibnamefont{Algarni}},
  \bibinfo{author}{\bibfnamefont{J.}~\bibnamefont{Partridge}},
  \bibinfo{author}{\bibfnamefont{L.}~\bibnamefont{Farrar}},
  \bibinfo{author}{\bibfnamefont{J.}~\bibnamefont{Zhou}}, \bibnamefont{et~al.},
  \bibinfo{journal}{arXiv preprint arXiv:2109.12588}  (\bibinfo{year}{2021}).

\end{thebibliography}

\end{document}